\documentclass[prb,superscriptaddress,floatfix,twocolumn,showpacs,amsmath,amssymb,final]{revtex4}
\usepackage{graphicx}
\usepackage{dcolumn}
\usepackage{bm}

\begin{document}
\title{Single-hole properties in the two-dimensional $t$-$J$ model on the honeycomb lattice}
\author{Andreas L\"uscher}
\affiliation{School of Physics, University of New South Wales, Sydney 2052, Australia}
\author{Andreas L\"auchli}
\affiliation{Institut Romand de Recherche Num\'erique en Physique des Mat\'eriaux (IRRMA), EPFL, CH-1015 Lausanne}
\author{Weihong Zheng}
\affiliation{School of Physics, University of New South Wales, Sydney 2052, Australia}
\author{Oleg P. Sushkov}
\affiliation{School of Physics, University of New South Wales, Sydney 2052, Australia}

\date{\today}
\begin{abstract} Motivated by the possible charge ordering in the recently discovered superconductor Na$_x$CoO$_2$~yH$_2$O,  which at filling $x=1/3$ would correspond to a half-filled honeycomb lattice, we investigate the single-hole dynamics of the $t$-$J$ model on this lattice. Using exact diagonalization, series expansion and the self-consistent Born approximation, we calculate the quasi-particle dispersion, bandwidth and residues and compare our findings with the well established results for the square lattice. Given the similarities between both lattices, it is not surprising that we find qualitatively similar features for the honeycomb lattice, namely an almost flat band along the edge of the magnetic Brillouin zone and well defined hole pockets around the corners. However,  we report a considerable disagreement between the three methods used concerning the bandwidth and discuss  possible origins of this discrepancy.
\end{abstract}


\maketitle

\section{Introduction}
The study of doped antiferromagnets has been a central subject in the area of strongly correlated electron systems. The pioneering work by Bulaevskii, Nagaev and Khomskii~\cite{bulaevskii67} as well as Brinkman and Rice~\cite{brinkman70} led the way to innumerable works, especially motivated by the discovery of high-temperature superconductors~\cite{bednorz86} and the suggestion by Anderson~\cite{anderson87} that the physics of these materials could be described by a two-dimensional Hubbard model on the square lattice with large on-site Coulomb repulsion. In this limit, the Hubbard model can be mapped to the $t$-$J$ model~\cite{hirsch85}, which has been shown to describe experimentally determined single-hole properties very accurately, considering $t$, $t'$, $t''$ and $J$ as independent variables~\cite{nazarenko95,leung97,belinicher96,sushkov97}. At half-filling, the $t$-$J$ model reduces to a Heisenberg antiferromagnet, because of the single occupancy restriction. The ground state of the Heisenberg model on a bipartite lattice is the quantum Ne\'el state with long-range magnetic order. Doping away from half-filling allows the charge carriers to propagate and hence scramble the N\'eel order, leaving strings of overturned spins behind. A variety of tools has been employed to study the hole motion in an antiferromagnetic background~\cite{dagotto94}, e.g., exact diagonalizations (ED)~\cite{barnes89,barnes92,poilblanc93,leung95,beran96,eder96,lee97}, the self-consistent Born approximation (SCBA)~\cite{schmittrink88,kane89,martinez91,liu92}, quantum Monte-Carlo simulations~\cite{duffy97,brunner00,mishchenko01} and series expansion~\cite{hamer98}. In the parameter range relevant for the copper-oxide materials, i.e., $J/t \approx 0.3$, these methods agree reasonably well~\cite{brunner00} and the main features emerging are a minimum of the quasi-particle dispersion at $\left(\frac{\pi}{2},\frac{\pi}{2}\right)$, with a very flat band along the edges of the Brillouin zone and a practically vanishing  quasi-particle residue at $\left(\pi,\pi\right)$.

The recent discovery of superconductivity in Na$_x$CoO$_2$~yH$_2$O~\cite{takada03} has heated interest in models on the triangular lattice, to which the SCBA has already been applied~\cite{azzouz96,trumper04}. In this material, the two-dimensional CoO$_2$ planes are separated by insulating layers of Na ions and water molecules. The cobalt ions at the center of the CO$_2$ octahedra form a triangular lattice. Superconductivity has been observed for doping $0.22\leq  x\leq0.47$~\cite{schaak03,chen04,milne04}. Spin and charge ordering tendencies in Na$_x$CoO$_2$ have been addressed recently~\cite{terasaki03,lee04,zheng04,watanabe05}, and could, at filling $x=1/3$, lead to an effectively half-filled honeycomb lattice, where hole and electron doping is possible. Estimations for the parameters in the $t$-$J$ model converge at around $J/t \approx 0.1-0.2$~\cite{baskaran03}, slightly smaller than for the copper oxides.

Motivated by the possible application to this new superconductor, we investigate the dynamics of a single hole on the honeycomb lattice, using exact diagonalization, series expansion and the SCBA and compare our findings with well established results for the square lattice. The rest of the paper is organized as follows. In Sec.~\ref{sec:model} we define the model and introduce the notations for the lattice geometries under consideration. Sec.~\ref{sec:methods} then contains a summary of the most important features of the applied methods. We first consider the simplified $t$-$J^z$ model in Sec.~\ref{sec:tJz}, with Ising rather than Heisenberg spin interactions where quantum fluctuations are absent and present our main results for the full $t$-$J$ model in Sec.~\ref{sec:tJ}. Finally, Sec.~\ref{sec:conclusion} contains our conclusions.

\section{Model\label{sec:model}}
The $t$-$J$ model originates from an effective low-energy description of the Hubbard model in the limit of strong Coulomb repulsion ($U \gg t$), where double occupancy is forbidden and $J=4t^2/U$. An accurate treatment of the large-$U$ limit  of the Hubbard model~\cite{hirsch85} leads to correlated hopping terms, which we neglect in this work, in order to simplify the discussion. The Hamiltonian therefore reads
\begin{multline}\label{eq:hamtJ}
H_\text{$t$-$J$} = -t \sum_{<i,j>,\ \sigma} {P\left(c_{i\sigma}^\dag c_{j\sigma} + H.c.\right)P} \\
+ J \sum_{<i,j>}\left( {\bf S}_i \cdot {\bf S}_j -\frac{1}{4}n_i n_j\right) \ .
\end{multline}
Here $c_{i\sigma}^\dag$ are the usual electron creation operators, with $n_i=c^\dag_{i\uparrow} c_{i\uparrow}+c^\dag_{i\downarrow}c_{i\downarrow}$, ${\bf S}_i$ are the electron spin operators and
$P$ the projection operator that excludes double occupancy. The square (honeycomb) lattice has coordination number $Z=4$ ($Z=3$). To simplify comparison between exact diagonalization, series expansion and the SCBA, we assign the zero energy level to the state with one static hole
\begin{equation} \label{eq:zeroenergy}
E_0=E_0^\text{Heisenberg}+\zeta ZJ/2=0 \ .
\end{equation}
Here $\zeta$ takes into account the quantum fluctuations. According to series expansion, $\zeta=1.08$ for the square lattice~\cite{hamer98}, and $\zeta=1.10$ for the honeycomb lattice,
whereas according to exact diagonalizations on the honeycomb lattice, we find $\zeta\approx 1.182$ for $N=18$ and 1.165 for $N=24$.

In order to gain additional understanding of the accuracy of the applied methods and the mechanism of hole propagation, it is insightful to study the simplified $t$-$J^z$ model
\begin{multline} \label{eq:hamtJz}
H_\text{$t$-$J^z$} = -t\sum_{<i,j>,\ \sigma} {P\left(c_{i\sigma}^\dag c_{j\sigma} + H.c.\right)P} \\
+ J ^z\sum_{<i,j>}\left( S_i^z \ S_j^z -\frac{1}{4}n_i n_j\right) \ ,
\end{multline}
with Ising rather than Heisenberg spin interactions. In the absence of spin fluctuations, $\zeta=1$ in Eq.~(\ref{eq:zeroenergy}).

\begin{figure}
\includegraphics[width=0.4\textwidth,clip]{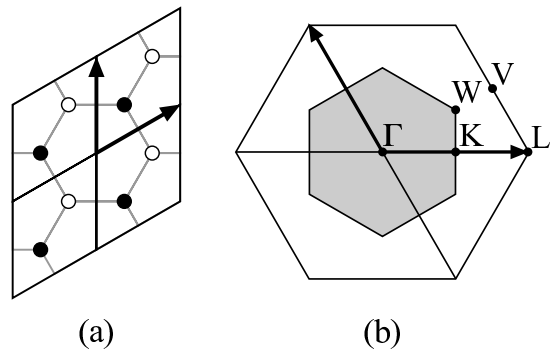}
\caption{\emph{Honeycomb lattice (described as a triangular lattice with two sites per unit cell) in real (a) and reciprocal space (b). The magnetic (shaded) and the full Brillouin zone coincide. The labels are defined in the text. Note that in contrast to the square lattice, the white area does not correspond to the Brillouin zone, but to the periodicity of the quasi-particle weight measured in ARPES experiments.} \label{fig:honeycomblattice}}
\end{figure}

\begin{figure}
\includegraphics[width=0.4\textwidth,clip]{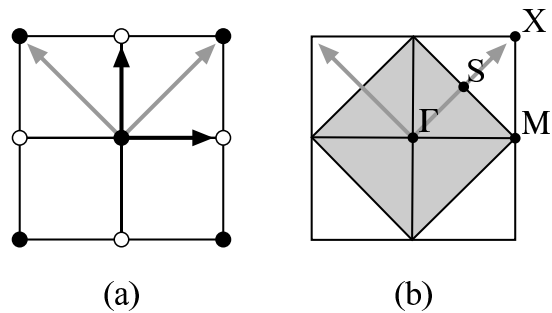}
\caption{\emph{Square lattice in real (a) and reciprocal space (b). The shaded area corresponds to the magnetic Brillouin zone. Highly symmetric points are mentioned in the text.}
\label{fig:squarelattice}}
\end{figure}

We study the above models on the square and the honeycomb lattice. The latter one being not a Bravais lattice, is best described as a triangular lattice with two sites per unit cell. The reciprocal lattice is then again triangular, and the first Brillouin zone is a honeycomb, as shown in Fig.~\ref{fig:honeycomblattice}.
Highly symmetric points are $\Gamma=\left(0,0\right)$, $K=\left(\frac{2\pi}{3},0\right)$, $L=\left(\frac{4\pi}{3},0\right)$, $V=\left(\pi,\frac{\pi}{\sqrt{3}}\right)$, and $W=\left(\frac{2\pi}{3},\frac{2\pi}{3\sqrt{3}}\right)$. Similar drawings for the square lattice are shown in Fig.~\ref{fig:squarelattice}, with the main purpose to introduce the notation for points of special interest $\Gamma=\left(0,0\right)$, $M=\left(\pi,0\right)$, $X=\left(\pi,\pi\right)$, and $S=\left(\frac{\pi}{2},\frac{\pi}{2}\right)$.

\section{Methods \label{sec:methods}}
We calculate several single-hole properties, namely the quasi-particle dispersion, the bandwidth and the residues, using exact diagonalization and series expansion methods as well as the SCBA. We briefly recall the ideas, advantages and possible difficulties associated with these three methods.

\subsection{Exact diagonalization}
In exact diagonalization one can rather easily calculate spectral functions directly in real frequency, using the so called continued-fraction technique \cite{gagliano87,dagotto94}. This  allows to obtain unbiased results, including residues, although only on finite-size samples. In special cases it is possible to calculate single-hole spectral functions for $t$-$J$ models on samples  up to $32$ sites, depending on the lattice symmetries. This method has been applied on numerous occasions to obtain single hole spectral functions, especially for the square lattice \cite{barnes89,barnes92,poilblanc93,leung95,beran96,eder96,lee97}. To our knowledge single-hole properties for the honeycomb lattice have not been studied in ED before. In the following we will mainly present results obtained on $N=18$ and $N=24$ sites samples. They share the advantage of having the symmetries of the infinite lattice and furthermore have the $W$ point in their discrete Brillouin zone, therefore giving a reasonable  estimate of the bandwidth. For some quantities we considered systems up to $N=32$ with one hole.

\subsection{Series expansions \label{sec:series}}
The Ising series expansion for the single-hole dispersion of the $t$-$J$ model has been done previously for the square lattice~\cite{hamer98}. To perform the series expansion, one needs to introduce an expansion parameter $x$, and divide the Hamiltonian into an unperturbed part ($H_0$) and a perturbation ($V$) as follows
\begin{equation*}
H/J = H_0 + x V \ ,
\end{equation*}
with
\begin{align*}
H_0 &= \sum_{\langle ij\rangle} \left(S_i^z S_j^z - \frac{1}{4} n_i n_j \right) + r \sum_i (-1)^i S_i^z \ ,\\
V &= - \frac{t}{J} \sum_{\langle ij\rangle ,\sigma} P \left(c^{\dag}_{i,\sigma} c_{j,\sigma} + c^{\dag}_{j,\sigma} c_{i,\sigma} \right) P \nonumber \\
&\hspace*{1cm} + \sum_{\langle ij\rangle} \left(S_i^x S_j^x + S_i^y S_j^y\right) - r \sum_i (-1)^i S_i^z \ ,
\end{align*}
with a staggered $z$-field of strength $r$ to improve convergence. This term vanishes in the total Hamiltonian in the isotropic exchange limit $x\to 1$. The Ising part of the exchange, $H_0$, forms the unperturbed Hamiltonian while the perturbation $V$ includes both the hole hopping terms and transverse spin fluctuations. The original Hamiltonian for the $t$-$J$ model is recovered in the limit $x=1$. We expect the series expansion to give more reliable results in the small $t/J$ region, where the perturbation $V$ is less important. The series for the single-hole dispersion has been computed to order $x^{13}$ for both the square and the honeycomb lattice, involving a list of 325737 and 28811 clusters with up to 13 sites, respectively. This extends the previous series for square lattice~\cite{hamer98} by only 2 terms. We also compute the new series for the quasi-particle residue, again to order $x^{13}$.

For the $t$-$J^z$ model, one can also form a direct expansion in powers of $t/J^z$. This series has been computed to order $(t/J^z)^{28}$ $\left[(t/J^z)^{26}\right]$ for square [honeycomb] lattice requiring a list of 4654284 [80905] clusters with up to 15 [14] sites. Previous results for the $t$-$J^z$ model on the square lattice~\cite{hamer98} used expansions of order $(t/J^z)^{21}$.

These series are available from the authors upon request.

\subsection{Self-consistent Born approximation}
From a general point of view, the SCBA works well for a collinear state, whenever vertex corrections are small. In the case of the hole-spin-wave vertex, single loop corrections are zero, and higher order corrections are small numerically~\cite{liu92}. Despite its simplicity, the SCBA has the disadvantage that its predictions are not immediately comparable to other methods or experimental results, because it distinguishes pseudo-spin, i.e., holes living on a particular sublattice, rather than spin variables.

\subsubsection{Introduction}
Let us start this reminder by defining the bare hole operators $d_i$, so that $d_{i}^{\dag} \propto c_{i\uparrow}$ on the $\uparrow$-sublattice ($A$) and  $d_{i}^\dag \propto c_{i \downarrow}$  on the $\downarrow$-sublattice ($B$). In momentum representation
\begin{align} \label{eq:d}
d_{{\bf k},A}^{\dag} &=\sqrt{\frac{2}{N (1/2+m)}}\sum_{i \in A} c_{i\uparrow}e^{i{\bf k}{\bf r}_i} \ , \nonumber \\
d_{{\bf k},B}^{\dag} &=\sqrt{\frac{2}{N (1/2+m)}}\sum_{j \in B} c_{j\downarrow}e^{i{\bf k}{\bf r}_j} \ .
\end{align}
$N$ is number of sites and $m= |\langle 0|S_i^z|0\rangle|$ the average sublattice magnetization, with $m \approx 0.3$ for the square lattice and $m \approx 0.24$ for the honeycomb lattice. We denote  the ground state of the undoped antiferromagnet, i.e. the quantum N\'eel state by $\left|0\right\rangle$. In the Ising case, there is no reduction of the sublattice magnetization due to quantum fluctuations and therefore $m=0.5$. In this case, $\left|0\right\rangle$ is the classical N\'eel state. The quasi-momentum ${\bf k}$ is limited to the magnetic Brillouin zone and the coefficients in~(\ref{eq:d}) provide the correct normalization
\begin{multline*} \label{eq:normd}
\langle 0|d_{{\bf k},A} d_{{\bf k},A}^{\dag}|0\rangle =\frac{2}{N (1/2+m)}\sum_{i \in
A} \langle 0|c_{i\uparrow}^{\dag}c_{i\uparrow}|0\rangle \\
=\frac{1}{1/2+m}\langle 0|\frac{1}{2}+S_i^z|0\rangle=1 \ .
\end{multline*}
The retarded hole Green's function is defined as
\begin{equation} \label{eq:greend}
G_{d,A}(\epsilon,{\bf k})=-i\int_0^\infty \langle 0| d_{{\bf k},A}(\tau) d_{{\bf k},A}^{\dag}(0)|0
\rangle e^{i \epsilon \tau} d\tau \ ,
\end{equation}
with the corresponding spectral function
\begin{equation*}
A_{d,A}(\epsilon,{\bf k})=-\frac{1}{\pi} \text{Im } G_{d,A}(\epsilon,{\bf k}) \ ,
\end{equation*}
and similar for $G_{d,B}(\epsilon,{\bf k})$.

For spin excitations, the usual linear spin-wave theory is used~\cite{manousakis91}. It is convenient to have two types of spin waves, $\alpha_{\bf q}^\dag$ with  $S^z=-1$, and $\beta_{\bf q}^\dag$ with $S^z=+1$, and ${\bf q}$ restricted to the magnetic Brillouin zone. The spin operators can then be expressed as
\begin{align*}
S_{i\in A}^+ &= 2 \sqrt{\frac{S}{N}} \sum_{\bf q}{e^{i {\bf q}\cdot {\bf r}_i}
\left(u_{\bf q} \alpha_{\bf q} + v_{\bf q} \beta_{-{\bf q}}^\dag\right)} \ , \\
S_{j\in B}^-&= 2 \sqrt{\frac{S}{N}} \sum_{\bf q} {e^{-i {\bf q}\cdot {\bf r}_j}
\left(u_{\bf q} \beta_{-{\bf q}}+v_{\bf q} \alpha_{\bf q}^\dag \right)} \ .
\end{align*}
Although spin-wave theory is an expansion in powers of $1/S$, its application to the $t$-$J$ model is restricted to the case $S=1/2$, because only in this limit is it possible to flip a spin by absorption or emission of a \emph{single} magnon. We therefore set $S=\frac{1}{2}$ from now on.

Since the honeycomb lattice lacks inversion symmetry, $\gamma_{\bf q}$ and the parameters of the Bogoliubov transformation are complex:
\begin{align*}
\omega_{\bf q} &= \sqrt{1-|\gamma_{\bf q}|^2} \ ,
\quad \gamma_{\bm q} = \frac{1}{Z}\sum_{\bm \rho} e^{i{\bf q} \cdot{\bm \rho}}= |\gamma_{\bf q}|e^{i \delta_{\bf
q}} \ , \\
u_{\bf q} &= \sqrt{\frac{1}{2}}\left(\frac{1}{\omega_{\bf q}}-1\right)^{1/2} \ ,\\
v_{\bf q} &= -\sqrt{\frac{1}{2}}\left(\frac{1}{\omega_{\bf q}}-1\right)^{1/2} e^{-i\delta_{\bf
q}} \ ,
\end{align*}
where ${\bm \rho}$ is the vector to nearest neighbors and $\delta_{\bf q}$ is the phase of $\gamma_{\bf q}$. In this notation, the spin-wave dispersion is equal to $\epsilon_{\bf q}=J Z /2 \, \omega_{\bf q}$. Hopping to a nearest neighbor in the Hamiltonian~(\ref{eq:hamtJ}) gives an interaction of the hole with spin waves
\begin{widetext}\begin{equation} \label{eq:hsw}
H_{h,sw}=\sum_{\bf k,q} \left( g_{\bf k,q} d_{{\bf k+q},A}^\dag d_{{\bf k},B} \alpha_{\bf q} +
g_{\bf k,q}^* d_{{\bf k+q},B}^{\dag} d_{{\bf k},A} \beta_{\bf q} + H.c.\right) \ ,
\end{equation}
with
\begin{align*}
g_{\bf k,q} &= \langle 0|\alpha_{\bf q}d_{{\bf k},B}|H_t| d_{{\bf k+q}, A}^{\dag}|0 \rangle
\nonumber \\
&=\frac{2}{N (1/2+m)} \langle 0|\alpha_{\bf q}\sum_{\langle i \in A, j \in B \rangle} c_{j \downarrow}^{\dag} e^{-i{\bf k}{\bf r}_j} \left(-t c^{\dag}_{i\sigma} c_{j\sigma}\right) c_{i\uparrow}e^{i({\bf k+q}){\bf r}_i}|0\rangle \\
&=\frac{2t}{N (1/2+m)} \sum_{\langle i \in A, j \in B \rangle}e^{-i{\bf k} {\bf r}_j+i({\bf
k+q}){\bf r}_i} \left( \langle 0|\alpha_{\bf q}c_{j \downarrow}^{\dag}c_{j \uparrow}
c_{i \uparrow}^{\dag}c_{i \uparrow}|0\rangle+ \langle 0|\alpha_{\bf q}c_{j \downarrow}^{\dag}c_{j
\downarrow} c_{i \downarrow}^{\dag}c_{i \uparrow}|0\rangle\right) \nonumber \\
&\approx\frac{2t}{N} \sum_{\langle i \in A, j \in B \rangle}e^{-i{\bf k}{\bf r}_j + i({\bf
k+q}){\bf r}_i} \langle 0|\alpha_{\bf q}\left(S_j^- +S_i^-\right)|0\rangle=
Zt\sqrt{\frac{2}{N}}(\gamma^*_{\bf k}u^*_{\bf q}+ \gamma^*_{\bf k+q}v_{\bf q}) \ ,
\end{align*} \end{widetext}
using the usual mean-field ground state factorization approximation
\begin{align*}
\langle 0|\alpha_{\bf q}c_{j \downarrow}^{\dag}c_{j \uparrow} c_{i \uparrow}^{\dag}c_{i \uparrow}
 |0\rangle &\approx \langle 0|\alpha_{\bf q}c_{j \downarrow}^{\dag}c_{j \uparrow}|0\rangle
\langle 0|c_{i \uparrow}^{\dag}c_{i \uparrow}|0\rangle \\
&= \langle 0|\alpha_{\bf q}S_j^-|0\rangle (1/2+m) \ .
\end{align*}

For a collinear state, the interaction~(\ref{eq:hsw}) forbids single loop corrections to the hole-spin-wave vertex and it has been shown that two and higher order loop corrections, which correspond to Trugman processes~\cite{trugman88} are small~\cite{martinez91,liu92}. This justifies the SCBA for $J_{min} \le J/t \ll 1$ according to which the hole Green's function satisfies the Dyson equation
\begin{equation}
\label{eq:dyson}
G_d(\epsilon, {\bf k})=\left(\epsilon -\sum_{\bf q}\left|g_{\bf k-q,q}\right|^2 G_d(\epsilon -\epsilon_{\bf
q},{\bf k-q}) + i 0 \right)^{-1}.
\end{equation}
Clearly, there is a $J_{min}$ below which the ground state of the system is found in the sector $\left|S_\text{tot}^z\right|> 1$, because in the limit $J/t \rightarrow 0$ the Nagaoka theorem applies, predicting a completely polarized ground state. A  DMRG calculation~\cite{white01} for the square lattice with up to $9\times 9$ sites estimates $J_{min}/t \approx 0.02-0.03$.

Due to the definition of the operators~(\ref{eq:d}), the Green's function~(\ref{eq:greend}) is invariant under translation with the inverse vector of the magnetic sublattice ${\bf Q}$.
\begin{equation*}
G_d(\epsilon,{\bf k+Q})=G_d(\epsilon,{\bf k}) \ .
\end{equation*}
For the square lattice ${\bf Q}=(\pm \pi,\pm \pi)$, and for the honeycomb lattice ${\bf Q}=(\pm
\frac{2\pi}{3},\pm \frac{2\pi}{\sqrt{3}})$.

\subsubsection{SCBA and ARPES experiments}
\begin{figure}
\includegraphics[width=0.4\textwidth,clip]{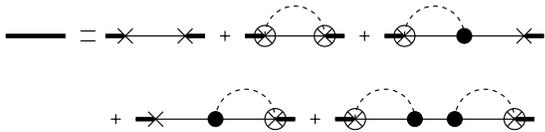}
\caption{\emph{Dyson equation relating the hole Green's functions $G_{c}$ (bold lines) and $G_{d}$ (solid lines). Dashed lines represent spin waves.} 
\label{fig:dyson}}
\end{figure}

As explained in detail in Ref.~\onlinecite{sushkov97}, the imaginary part of the hole Green's function $G_{d}$  is not directly measurable in ARPES experiments, because the incoming photon does not distinguish the sublattices. For comparison with experiments, we define an annihilation operator with given spin $\sigma$
\begin{equation} \label{eq:c}
c_{{\bf k},\sigma} =\sqrt{\frac{2}{N}}\sum_i c_{i \sigma}e^{i{\bf k}{\bf r}_i} \ .
\end{equation}
The summation extends over all $N$ sites and ${\bf k}$ is the momentum of the free photoelectron detected in the experiment, i.e., it is not restricted to the Brillouin zone. The normalization is chosen to fulfill $\langle 0|c^{\dag}_{{\bf k} \sigma} c_{{\bf k} \sigma}|0\rangle= \frac{2}{N} \langle 0|\sum_i c^{\dag}_{i\sigma}c_{i\sigma}  |0\rangle=1$.
In this high-energy approximation, every site is regarded as the possible origin of the emission of a free photoelectron. For the square lattice, this definition coincides with the normal Fourier transformation, but for the honeycomb lattice, this is not the case, since the full lattice is not a Bravais lattice. Eq.~(\ref{eq:c}) can therefore not be inverted in general. The spectral function measured in ARPES corresponds to the imaginary part of the retarded Green's function
\begin{equation} \label{eq:greenc}
G_c(\epsilon,{\bf k})=-i\int_0^\infty \langle 0|c_{{\bf k},\sigma}^{\dag} (\tau)c_{{\bf k},\sigma}(0)|0 \rangle e^{i \epsilon \tau} d\tau.
\end{equation}
\begin{figure}
\includegraphics[width=0.4\textwidth,clip]{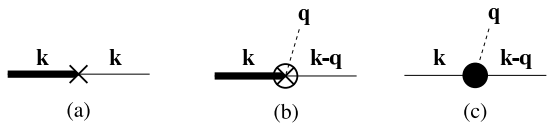}
\caption{\emph{Vertices representing a single hole creation (a), a hole plus spin-wave creation (b) and the usual hole-spin-wave interaction (c).
 Bold lines correspond to $G_{c}$, solid lines represent $G_{d}$, and dashed lines are spin waves.}
\label{fig:vertices}}
\end{figure}
The operator $c_{{\bf k},\downarrow}$ acting on the quantum N\'eel state can either produce a single hole state, by removing an electron with spin $\sigma=\downarrow$ from sublattice $B$, or it can produce a hole plus spin-wave state, by removing such an electron from sublattice $A$. We adopt the notation of Ref.~\onlinecite{sushkov97} and denote the corresponding amplitudes by  $a_{\bf k}$ and   $b_{\bf k,q}$, respectively
 \begin{align*}
a_{\bf k} &=\langle 0| d_{{\bf k},B} c_{{\bf k},\downarrow}|0\rangle =
\sqrt{1/2+m} \\ \label{eq:bk}
b_{\bf k,q}&=\langle 0|\beta_{\bf q} d_{{\bf k-q},A} c_{{\bf k},\downarrow}|0\rangle\\ 
&= \sqrt{ \frac{2}{N\left(1/2+m\right)}}  v_{\bf q}^* \approx \sqrt{\frac{2}{N}} v_{\bf q}^*.
\end{align*}
These vertices are shown in Fig.~\ref{fig:vertices} and afterwards are used in Fig.~\ref{fig:dyson} to derive the Dyson equation for the Green's function of a hole with fixed spin $G_{c}$.
In analytical form, the Green's function $G_c$~(\ref{eq:greenc}) is given by
\begin{widetext}\begin{equation}\begin{split} \label{eq:dyson3}
G_c(\epsilon,{\bf k})=a_{\bf k}^2 G_d(\epsilon,{\bf k}) +\sum_{\bf q}\left|b_{\bf k,q}\right|^2  G_d(\epsilon-\epsilon_{\bf q},{\bf k-q})
+2 a_{\bf k} G_d(\epsilon,{\bf k})\left[ \sum_{\bf q}\text{Re} \left(b_{\bf k,q} g^*_{\bf k-q,q}\right) G_d(\epsilon-\epsilon_{\bf q},{\bf k-q}) \right] \\
+G_d(\epsilon,{\bf k}) \left[ \sum_{\bf q}b_{\bf k,q} g^*_{\bf k-q,q} G_d(\epsilon-\epsilon_{\bf q},{\bf k-q}) \right] \left[ \sum_{\bf q}b^*_{\bf k,q} g_{\bf k-q,q} G_d(\epsilon-\epsilon_{\bf q},{\bf k-q}) \right] \ .
\end{split} \end{equation} \end{widetext}
The numerical solution of~(\ref{eq:dyson}) is straightforward. We replace $i0$ by $0.05 i$ and use $32 \times 32$ points in the Brillouin zone and an energy interval $\Delta\epsilon \leq 0.002$ to carry out the iterations. Convergence is reached after around 50 iterations for small values of $J/t$. $G_{c}$ is subsequently calculated according to Eq.~(\ref{eq:dyson3}).
\begin{figure}
\includegraphics[width=0.4\textwidth,clip]{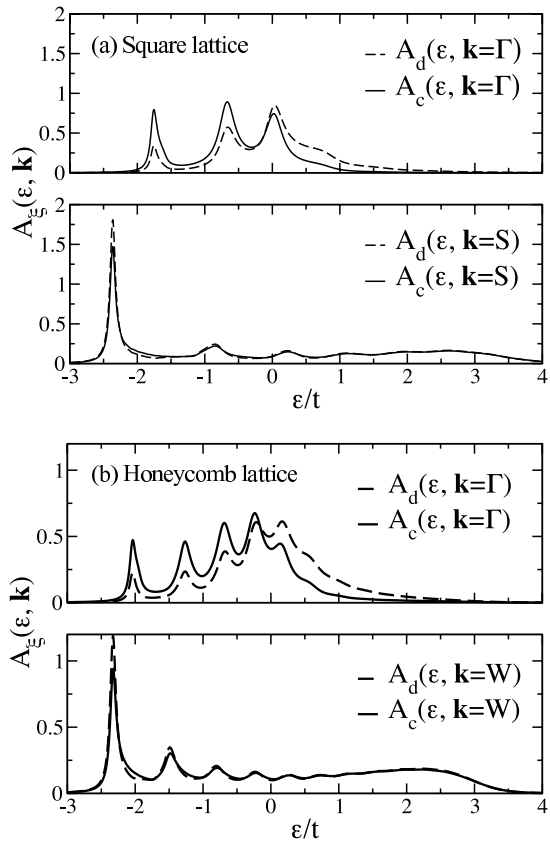}
\caption{\emph{(a) Spectral functions of a single hole in the $t$-$J$ model on the square lattice with $J/t=0.3$ at the minimum $S$ and the maximum $\Gamma$ of the dispersion. (b) Similar results for the honeycomb lattice with $J/t=0.2$ at the minimum $W$ and the maximum $\Gamma$ of the dispersion.}
\label{fig:spectralfunctions}}
\end{figure}
Figs.~\ref{fig:spectralfunctions}(a) and (b) show the spectral functions $A_{d}$ and $A_{c}$ for a hole on the square lattice with $J/t=0.3$, and on the honeycomb lattice with $J/t$=0.2, respectively. A hole with fixed spin $G_{c}$ has the same dispersion as a hole linked to a particular sublattice $G_{d}$, but clearly, the areas under the peaks is different. A discussion of the origin of these peaks and a comparison with ED calculations is presented in Sec.~\ref{sec:tJH}. At this point, it seems useful to comment on the physical difference between the two Green's functions.  The electron creation operator $c^\dag_{{\bf k},\sigma}$ does not correspond to a true quasi-particle of the \emph{system with long-range antiferromagnetic order}, because the spin $\sigma$ is not conserved. However, the hole creation operator associated with a particular sublattice $d^\dag_{{\bf k},A/B}$ is well defined in this case and the quasi-particle residue corresponds to the area under the peak of the spectral function. In contrast, exact diagonalization and series expansion do not rely on the existence of long-range order and give direct access to the Green's function $G_{c}$. While there is no ambiguity concerning the single-hole dispersion, one has to be careful comparing residues, which we define as
\begin{equation*}
Z_\xi\left({\bf k}\right)=  \int_{-\infty}^{\epsilon_p} A_\xi\left(\epsilon,{\bf k}\right) d\epsilon ,
\end{equation*}
where $\epsilon_p$ is the cut-off chosen to include only the lowest-lying peak. This definition has the disadvantage that $\epsilon_p$ is not well defined and that the tails of higher excitations are also taken into account. To circumvent these problems, we make the assumption of Lorentzian peaks and easily deduce the residue. Let us finally mention the finite-size effects that come into play in the calculation of  $G_{c}$. While for the hole Green's function with fixed pseudo-spin, $G_{d}$, grids with more than $32 \times 32$ points in the Brillouin zone yield essentially the same results and the energy grid with $\Delta\epsilon=0.002$ fulfills the sum rule with good accuracy, i.e. the error is smaller than 0.5\%. As can be deduced from the sum rules~\cite{sushkov97}, the situation is more subtle in the case of $G_{c}$
\begin{align*}
\int_{-\infty}^\infty A_d\left(\epsilon,{\bf k}\right) d\epsilon &=1 \ , \\
\int_{-\infty}^\infty A_c\left(\epsilon,{\bf k}\right) d\epsilon &=a_{\bf k}^2+\sum_{\bf q}
\left| b_{{\bf k}, {\bf q}} \right|^2=1 \ .
\end{align*}
Clearly, a grid of only $32 \times 32$ points in the Brillouin zone does not lead to $m=\frac{1}{2}-\frac{2}{N}\sum_{\bf q}  \left| v_{\bf q}\right|^2\approx 0.24$, and the error in the sum rule is of around 3\%. We solve this problem by choosing $m$ according to the grid, i.e., $m_{32} \approx 0.26$.

\subsubsection{$J/t \rightarrow 0$ limit}
Below $J_{min}/t$, the ground state is found in the sector $\left|S_{tot}^z\right| > 1$, which cannot be accessed within the SCBA. In other words, in this limit, the ``string'' picture is no longer valid and the hole
energy is minimized by a ferromagnetic ``spin bag'', in which the hole can hop without modifying the
background. For a circular bag with radius $R\gg a=1$, the hole energy is given by~\cite{shraiman88}
\begin{equation*}
E_{0}=-Z t+ \xi \sqrt{J t} \ ,
\end{equation*}
where $\xi=8.53$ and $\xi=5.61$ for the square and the honeycomb lattice, respectively. We would like to point out, that these considerations are only valid in the case of the $t$-$J$ model, because in the absence of quantum fluctuations, it is energetically favorable to align the spins in the plane perpendicular to the quantization axis~\cite{white01}. The single-hole ground state therefore carries a total spin $S_{tot}^z = 0$.

\subsubsection{SCBA for the $t$-$J^z$ model\label{sec:SCBAtJz}}
In the SCBA for the Ising case~(\ref{eq:hamtJz}), the Bogoliubov parameters and the dispersion are given by $u_{\bf q}=1$, $v_{\bf q}=0$, and $\epsilon_{\bf q}=J^zZ/2$. The Dyson equation~(\ref{eq:dyson}) therefore becomes momentum independent
\begin{equation}
\label{eq:dysontJz}
G_d\left(\epsilon\right) = \frac{1}{\epsilon-Z t^2 G_{d}\left(\epsilon - J^z Z/2\right)}
\end{equation}
and can be solved analytically~\cite{starykh95}
\begin{equation} \label{eq:GtJz}
G_{d}\left(\epsilon\right)=-\frac{1}{\sqrt{Z} t} \frac{J_{-2\epsilon/\left(J^z Z\right)}\left(\frac{4
t}{J^z \sqrt{Z}}\right)} {J_{-2\epsilon/\left(J^z Z\right)-1}\left(\frac{4 t}{J^z \sqrt{Z}}\right)} \ .
\end{equation}
Here $J_{\nu}\left(x\right)$ is a Bessel function of the first kind. Note that the solution~(\ref{eq:GtJz}) is not an exact solution of the $t$-$J^z$ model. The hole, dispersionless in the SCBA, is mobile even in the absence of quantum fluctuations due to Trugman processes~\cite{trugman88}. The simplest one corresponds to the hole motion around an elementary plaquette of the lattice and effectively leads to the propagation of the hole to a next-nearest neighbor without leaving overturned spins behind. These processes are of sixth and tenth order in $t/J^z$ for the square and the honeycomb lattice, respectively, and therefore negligible. More importantly, it was shown in Ref.~\onlinecite{chernyshev99} using a diagrammatic study, that the inclusion of higher order terms (non-linear hole-magnon, static hole-magnon, direct hole-hole and magnon-magnon) can lead to substantially different results. The intuitive picture emerging from that work is as follows: the dynamic hole-magnon interaction considered in the Dyson equation~(\ref{eq:dysontJz}) does not take into account the presence of the hole, which reduces the number of possible sites the hole can hop to and also the energy associated with a flipped spin. In a first step, the hole can hop to $Z$ nearest neighbors, leaving a flipped spin with energy $J^z (Z-1)/2$ behind. For the second hopping, there are only $Z-1$ possible targets and flipping a spin costs an energy $J^z (Z-2)/2$. The Dyson equation~(\ref{eq:dysontJz}) therefore has to be replaced by
\begin{align} \label{eq:dyson2}
G_{d}\left(\epsilon\right) &= \frac{1}{\epsilon-Z t^2 \tilde{G_{d}}\left(\epsilon -  J^z \left(Z-1\right)/2\right)}
\nonumber \\
\tilde{G_{d}}\left(\epsilon\right) &= \frac{1}{\epsilon-(Z-1) t^2 \tilde{G_{d}}\left(  \epsilon-
J^z \left( Z - 2 \right)/2 \right)}.
\end{align}
In this modified approach, the SCBA gives the exact result in the limit $J/t \rightarrow \infty$, where the ground state energy is equal to $E=- 2 Z t^2 / (J^z (Z-1)) $. For the full $t$-$J$ model, these geometrical considerations are believed to be less important, because quantum fluctuations restore overturned spins~\cite{martinez91}.

\section{Results for the $t$-$J^z$ model \label{sec:tJz}}
Studying the simplified $t$-$J^z$ model has the advantage that series expansion methods are perfectly suited for this task and that they can therefore serve as a benchmark for other methods. However, it is difficult to make predictions for the $t$-$J$ model on the basis of these results, since the modifications arising from quantum fluctuations are hard to quantify. Nevertheless, a comparison between the usual~(\ref{eq:dysontJz}) and modified~(\ref{eq:dyson2}) SCBA on the two lattices gives interesting insight in the importance of higher order terms discussed in Sec.~\ref{sec:SCBAtJz}, and a comparison with series expansion results allows to estimate the influence of Trugman processes. 
\begin{figure}
\includegraphics[width=0.4\textwidth,clip]{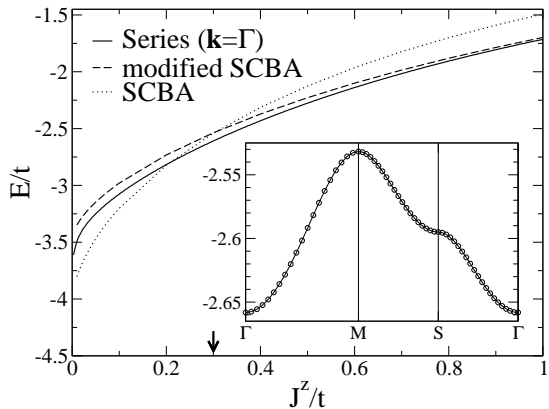}
\caption{\emph{Single-hole energy of the $t$-$J^z$ model on the square lattice. The SCBA results are
dispersionless and based on the usual and modified Dyson equations~(\ref{eq:dyson}) and (\ref{eq:dyson2}),respectively. The series expansion results are plotted for the minimum $\Gamma$ of the dispersion, which is shown in the inset for $J^z/t=0.3$.}
\label{fig:tJzSQen}}
\end{figure}
The common interpretation of the hole propagation in the $t$-$J^z$ model relies on the ``string picture'': as the hole moves in the rigid N\'eel state, it scrambles the antiferromagnetic order and leaves a  string of overturned spins behind. The cost in energy for these hoppings is thus proportional to the distance and the problem can be interpreted as a particle moving in a linear potential~\cite{bulaevskii67} and the hole energies then show the well known $\left(J^z/t\right)^{2/3}$ behavior. From a numerical solution of the modified Dyson equation~(\ref{eq:dyson2}), we find that the hole energies (of the ground state and the first few excited states) can be fitted by 
\begin{equation*}
E/t = -2\sqrt{Z-1} + \alpha \left(J^z/t\right) + \beta \left(J^z/t\right)^{2/3} \ ,
\end{equation*}
with coefficients $\alpha$ and $\beta$ depending on the lattice geometry and the state under consideration. An analytical estimation of $\beta$ can be found in Ref.~\onlinecite{chernyshev99}. 

\subsection{Square lattice\label{tJzSQ}}
Fig.~\ref{fig:tJzSQen} shows the well known results for the single-hole ground state energies of the $t$-$J^z$ model on the square lattice obtained via series expansion~\cite{hamer98} and SCBA~\cite{martinez91,liu92}.  Since the SCBA does not take into account closed-loop paths, its results are dispersionless and we compare it with the series expansion results at ${\bf k}=\Gamma$. The dispersion for $J^z/t=0.3$ is shown in the inset. It essentially represents hopping to next-nearest neighbors by loops turning $1.5$ times around elementary plaquettes, well described by $E({\bf k})=A\left( \cos(k_x+k_y) + \cos(k_x- k_y)\right)+C$. It is interesting to note that in contrast to the honeycomb lattice, the dispersion does not include hoppings to {\it all} next-nearest neighbors, because only sites on the same plaquette can be reached within a single loop. Next-nearest neighbors on adjacent plaquettes require two successive loops and are therefore less important. As can be seen in Fig.~\ref{fig:tJzSQen}, the modified Dyson equation~(\ref{eq:dyson2}) leads to a very accurate description of the hole. 
\begin{figure}
\includegraphics[width=0.4\textwidth,clip]{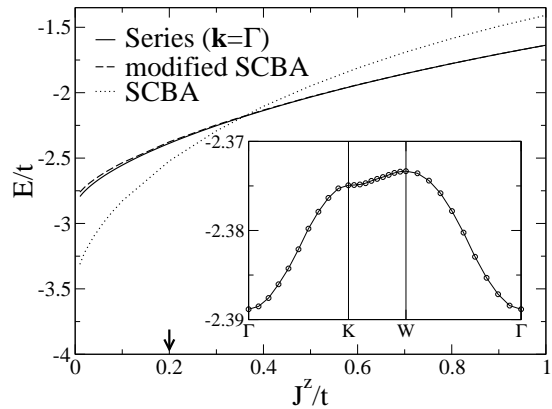}
\caption{\emph{Single-hole energy of the $t$-$J^z$ model on the honeycomb lattice. Trugman processes are less important than on the square lattice, as can be seen from the remarkable agreement between the modfied SCBA and series expansion results, and also from the narrow bandwidth of the dispersion shown in the inset for $J^z/t=0.2$.} \label{fig:tJzHen}}
\end{figure}

\subsection{Honeycomb lattice\label{tJzH}}
The single-hole energies of the $t$-$J^z$ model on the honeycomb lattice are presented in Fig.~\ref{fig:tJzHen}, with an inset showing the dispersion for $J^z/t=0.2$ obtained via series expansion. As expected, the bandwidth is much smaller than on the square lattice, because Trugman processes are of higher order and therefore less important. Similar to the square lattice, the dispersion represents hoppings to next-nearest neighbors and can therefore be described by $E({\bf k}) = A \left( 2 \cos\left(\sqrt{3} k_{y}\right) + 4 \cos\left(\sqrt{3}/2 k_{y}\right) \cos\left(3/2 k_{x}\right)\right) + C$.
\begin{figure*}
\includegraphics[width=0.9\textwidth,clip]{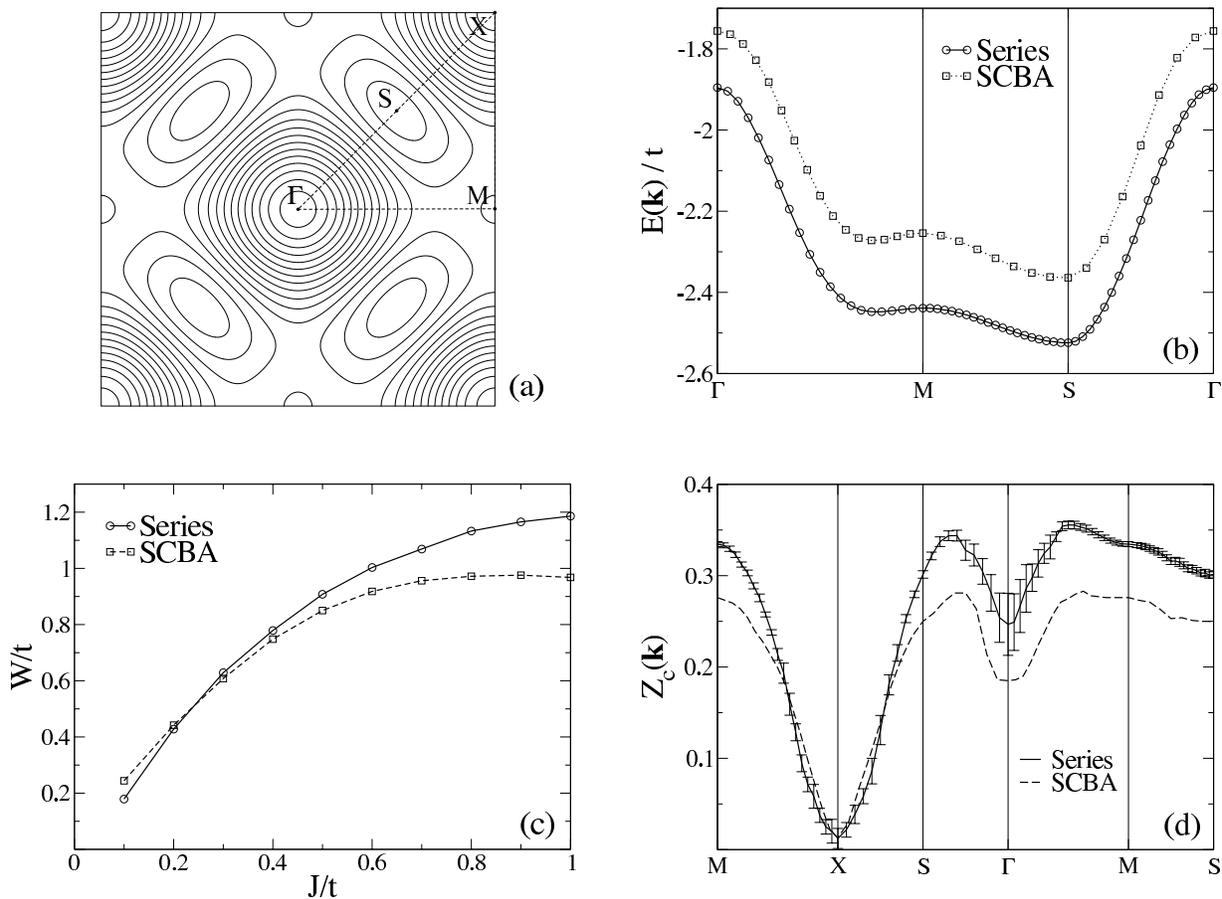}
\caption{\emph{Single-hole properties of the $t$-$J$ model on the square lattice: (a) Contour plot of the dispersion for $J/t=0.3$, with maximum at $\Gamma=(0,0)$ and minima (hole pockets) at $S=(\pm\frac{\pi}{2},\pm\frac{\pi}{2})$. (b) Dispersion for $J/t=0.3$ along the irreducible wedge of the Brillouin zone, see Fig.~\ref{fig:squarelattice} for notations. Apart from a global offset, the SCBA and series expansion results are in good agreement. (c) Bandwidth $W/t$ as a function of $J/t$. The agreement between series expansion and SCBA is very good up to $J/t\approx0.7$. (d) Quasi-particle residue $Z_c$ for $J/t=0.3$.}\label{fig:tJSQ}}
\end{figure*}
Because all next-nearest neighbors can be reached by one elementary loop, the dispersion corresponds to hopping inside a given sublattice. Results based on the modified Dyson equation~(\ref{eq:dyson2}) agree well with series expansion calculations. However, it is interesting to note that the corrections due to the presence of the hole are more important for the honeycomb than for the square lattice. For instance, for $J^z/t=0.1$ the difference for the honeycomb lattice is 12\%, but for the square lattice it is only 4\%. 

\section{Results for the $t$-$J$ model \label{sec:tJ}}
The $t$-$J$ model on the square lattice has been the object of numerous studies, involving many different methods. In order to compare our new findings for the honeycomb lattice with the well known calculations for the square lattice, we briefly review some of these previous results and in addition, present a new calculation of the quasi-particle residue using series expansion.

\subsection{Square lattice\label{sec:tJSQ}}
Fig.~\ref{fig:tJSQ}(a) shows the contour plot of the single-hole dispersion of the $t$-$J$ model on the square lattice for $J/t=0.3$. One clearly identifies four hole pockets centred at $S=(\pm\frac{\pi}{2},\pm\frac{\pi}{2})$, which are stretched along the face of the magnetic Brillouin zone. The maximum is found at  $\Gamma=(0,0)$. The dispersion along $\Gamma M S \Gamma$ is shown in Fig.~\ref{fig:tJSQ}(b). 
Apart from a global offset, series expansion and SCBA agree perfectly well. The main features, namely the dispersion minimum at $S$ and the flat band along the edge of the magnetic Brillouin zone already emerge by applying second order perturbation theory, and can therefore be described by hopping to next-nearest neighbors, i.e. hopping inside a given sublattice.  Above $J/t \approx 0.4$ the dispersion can thus be fitted by $E\left({\bf k}\right)= 2 A \left( \cos\left(k_{x}+k_{y}\right) + \cos\left(k_{x}-k_{y}\right)\right)+2 B \left( \cos\left(2 k_{x} \right) + \cos\left(2 k_{y}\right)\right)+C$.
 Compared to the $t$-$J^z$ model, where only hopping to next-nearest neighbors along the diagonals are relevant and hence $B=0$, the dispersion extrema are shifted from $M$ to $S$ and the overall shape changes substantially. For smaller values of $J/t$, the hole can hop further and additional hopping terms have to be taken into account. These contribution lead to more pronounced hole pockets. The agreement between the SCBA and series expansion is very good up to $J/t\approx0.7$, as can be seen in Fig.~\ref{fig:tJSQ}(c), showing the bandwidth as a function of $J/t$. The static magnon-hole interaction becomes more important with increasing $J/t$, and its neglection leads to the observed underestimation of the bandwidth in the SCBA. The residues $Z_c\left({\bf k}\right)$ for the $t$-$J$ model on the square lattice for $J/t=0.3$ are shown in Fig.~\ref{fig:tJSQ}(d). The main characteristic obtained via series expansion and SCBA is the almost complete suppression of the quasi-particle weight at $X$. 

\subsection{Honeyomb lattice\label{sec:tJH}}
\begin{figure}
\includegraphics[width=0.4\textwidth,clip]{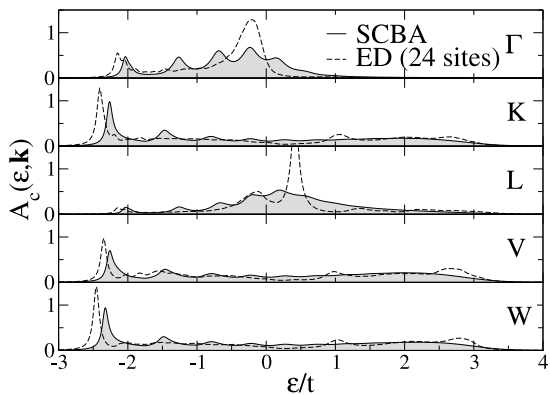}
\caption{\emph{Comparison between single-hole spectral functions on the honeycomb lattice obtained with the SCBA and exact diagonalization calculations for $J/t=0.2$.}
\label{fig:spectralfunctionsH}}
\end{figure}
Let us start this paragraph by examining the single-hole spectral functions for the honeycomb lattice shown in Fig.~\ref{fig:spectralfunctionsH}. For the square lattice, extensive discussions on this topic can be found in Refs.~\onlinecite{martinez91} and \onlinecite{liu92}. Throughout the hole Brillouin zone, there is a well-defined quasi-particle ground state. Apart from a global energy shift, ED and the SCBA lead to similar results concerning the weight and the shape of this lowest-lying peak. However, the structure of the excited states is different. The SCBA predicts several well pronounced quasi-particle excitations (which can be explained according to the ``string-picture'' developed for the $t$-$J^z$ model), whereas ED results show a rather featureless incoherent tail for ${\bf k}=K,L,W$ and a second dominant peak at much higher energies at ${\bf k}=\Gamma,L$. Given the simplicity of the SCBA, the overall agreement of the spectral functions is in our opinion quite remarkable, especially when focussing on ground state properties.

\begin{figure*}
\includegraphics[width=0.9\textwidth,clip]{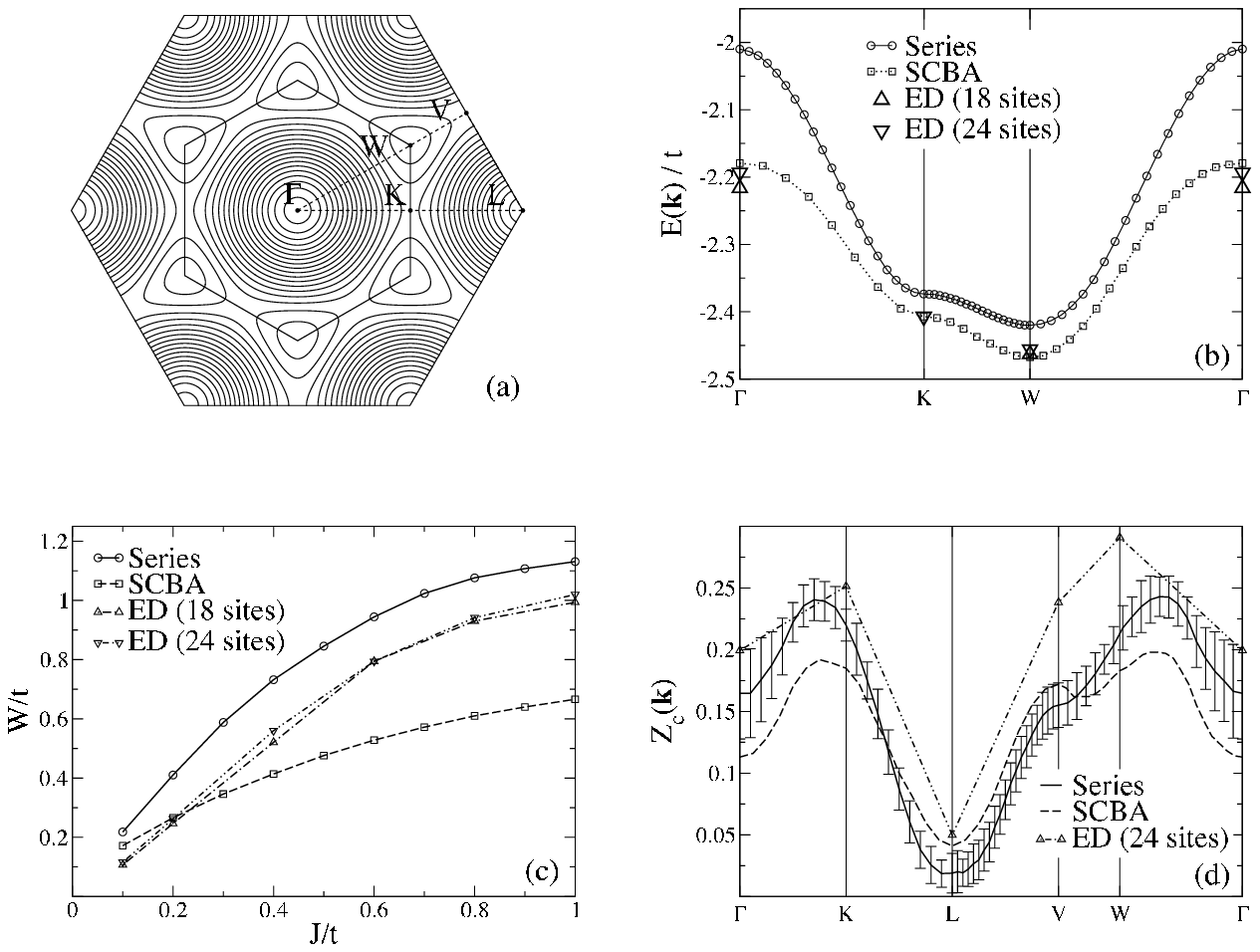}
\caption{\emph{Single-hole properties of the $t$-$J$ model on the honeycomb lattice: (a) Contour plot of dispersion for $J/t=0.2$. The maximum is found at $\Gamma=(0,0)$ and six circular hole pockets are located in the corners of the magnetic Brillouin zone $W=(\pm\frac{2\pi}{3},\pm\frac{2\pi}{3\sqrt{3}})$ and $W=(0,\pm\frac{4\pi}{3\sqrt{3}})$. (b) Dispersion for $J/t=0.2$ along the irreducible wedge of the Brillouin zone, see Fig.~\ref{fig:honeycomblattice} for notations. The overall shape of the dispersion is the same for all three methods, and essentially due to hopping to next-nearest neighbors, but series expansion predicts a substantially larger bandwidth. (c) Bandwidth $W/t$ as a function of $J/t$.The SCBA is valid only at small $J/t$, where it agrees well with ED. (d) Quasi-particle residue $Z_c$ for $J/t=0.2$.}\label{fig:tJH}}
\end{figure*}
Fig.~\ref{fig:tJH}(a) shows a contour plot of the single-hole dispersion for  $J/t=0.2$. The circular hole pockets are located in the corners of the magnetic Brillouin zone $W=(\pm\frac{2\pi}{3},\pm\frac{2\pi}{3\sqrt{3}})$ and $W=(0,\pm\frac{4\pi}{3\sqrt{3}})$, and the maximum is found at $\Gamma=(0,0)$. The dispersion along the irreducible wedge $\Gamma K W \Gamma$ is shown in Fig.~\ref{fig:tJH}(b). Similar to the square lattice, one finds a very flat band along the edge of the Brillouin zone. Above $J/t\approx0.7$, the dispersion is well described by next-nearest neighbor hopping $E({\bf k})=A \left( 2 \cos\left(\sqrt{3} k_{y}\right) + 4 \cos\left(\sqrt{3}/2 k_{y}\right) \cos\left(3/2 k_{x}\right)\right) + C$. Below  $J/t\approx0.7$ further hopping terms become increasingly important and lead to more pronounced hole pockets, analoguos to the square lattice. As can be seen in Fig.~\ref{fig:tJH}(b), the overall shape of the dispersion is very similar in ED calculations, the SCBA and series expansion. However, there is a discrepancy concerning the bandwidth: while ED (with up to 32 sites) and the SCBA lead to quantitatively similar results for $J/t\leq 0.3$, series expansion predicts a substantially larger bandwidth, see Fig.\ref{fig:tJH}(c).  From a general point of view, the validity of the SCBA is limited to small $J/t$, where the neglected hole-magnon interactions are small. The results for the $t$-$J^z$ model imply that these corrections are more important for the honeycomb lattice than for the square lattice, because of the reduced number of nearest neighbors. This implication is confirmed by the good agreement with ED calculations for $J/t\leq0.3$. Still, the noticeably larger bandwidth obtained by series expansion is quite puzzling. It seems that the difference is essentially due to the very steep band crossing the Brillouin zone from $\Gamma$ to $K$ and $W$, because the flat part along the edge from $K$ to $W$ practically coincides with ED and the SCBA. Interestingly, the three methods lead to almost identical results for the quasi-particle weight shown in Fig.~\ref{fig:tJH}(d). Analogous to the square lattice, the weight is substantially reduced but still finite at $L$. In view of the noticeably different predictions concerning the bandwidth, it would be interesting to compare these results with quantum Monte-Carlo calculations, which could elucidate the problem from a different point of view.

\section{Conclusion \label{sec:conclusion}}
For the first time, we have calculated single-hole properties of both the $t$-$J^z$ and the $t$-$J$ model on the honeycomb lattice using exact diagonalization, series expansion methods and the self-consistent Born approximation. We restrict our analysis to the physically relevant regime $J/t \leq 1$. Analogous to calculations for the very similar square lattice, we find an almost flat quasi-particle band along the edge of the magnetic Brillouin zone and well defined circular hole pockets in the corners. The spectral functions show well-defined quasi-particle peaks at the bottom of the spectrum throughout the whole Brillouin zone. We find a good agreement between the three methods applied concerning the shape of the single-hole dispersion and the quasi-particle weight. An analysis of the bandwidth clearly shows that the self-consistent Born approximation is quite accurate for $J/t \lesssim 0.3$, but leads to a substantial underestimation of the bandwidth for larger $J/t$. Series expansion on the other hand seems to overestimate the bandwidth in the small $J/t$ region. It is likely that this bandwidth disagreement is due to the reduced ``dimensionality'' of the honeycomb lattice, because the number of nearest neighbors ($Z=3$) lies between the square lattice ($Z=4$) and the linear chain ($Z=2$). This makes the problem especially interesting and a Monte-Carlo calculation would be very helpful to finalize the findings of the present work.

\acknowledgments{It is a pleasure to thank Jesko Sirker and Alexander Weisse for stimulating discussions and Alexander Chernyshev for helpful comments on the manuscript. The exact diagonalizations have been performed on the Altix SMP machine at EPFL. 
}


\begin{thebibliography}{99}

\bibitem{bulaevskii67} L.~N. Bulaevskii, E.~L. Nagaev, and D.~I. Khomskii, Sov. Phys. JETP {\bf 27}, 638 (1967)
\bibitem{brinkman70} W.~F. Brinkman and T.~M. Rice, \prb {\bf 2},  1324  (1970).
\bibitem{bednorz86} G.~J. Bednorz and K.~A. M\"uller, Z. Phys. B {\bf 64},  188  (1986).
\bibitem{anderson87} P.~W. Anderson, Science {\bf 235},  1196  (1987).

\bibitem{hirsch85} J.~E. Hirsch, \prl {\bf 54}, 1317 (1985).
\bibitem{nazarenko95} A. Nazarenko, K.~J.~E. Vos, S. Haas, E. Dagotto, and R.~J. Gooding, \prb {\bf 51}, R8676 (1995).
\bibitem{leung97} P.~W. Leung, B.~O. Wells, and R.~J. Gooding, \prb {\bf 56}, 6320 (1997).
\bibitem{belinicher96} V.~I. Belinicher, A.L. Chernyshev, and V.A. Shubin, \prb {\bf 54}, R14914 (1996).
\bibitem{sushkov97} O.~P. Sushkov, G.~A. Sawatzky, R. Eder, and H. Eskes, \prb {\bf 56}, 11769 (1997).

 \bibitem{dagotto94} For a review, see E. Dagotto, Rev. Mod. Phys. {\bf 66},  763  (1994).
\bibitem{barnes89} T. Barnes, E. Dagotto, A. Moreo, and E.~S. Swanson, \prb {\bf 40}, R10977(1989);
\bibitem{barnes92} T. Barnes, A.~E. Jacobs, M.~D. Kovarik, and W.~G. Macready, \prb {\bf 45}, 256(1992).
\bibitem{poilblanc93} D. Poilblanc, T. Ziman, H.~J. Schulz, and E. Dagotto, \prb {\bf 47}, 14267 (1993).
\bibitem{beran96} P. Beran, D. Poilblanc, and R.~B. Laughlin, Nucl. Phys. B {\bf 473}, 707 (1996).
\bibitem{leung95} P.~W. Leung and R.~J. Gooding, \prb {\bf 52}, R15711(1995).
\bibitem{eder96} R. Eder, Y. Ohta, and G.~A. Sawatzky, \prb {\bf 55}, R3414 (1996).
\bibitem{lee97} T.~K. Lee and C.~T. Shih, \prb {\bf 55}, R5983 (1997).
\bibitem{schmittrink88} S. Schmitt-Rink ,C.~M. Varma, A.~E. Ruckenstein, \prl {\bf 60} 2793 (1988).
\bibitem{kane89} C.~L. Kane, P.~A. Lee, and N. Read, \prb {\bf 39},  6880  (1989).
\bibitem{martinez91} G. Martinez and P. Horsch, \prb {\bf 44},  317  (1991).
\bibitem{liu92} Z. Liu and E. Manousakis, \prb {\bf 45},  2425  (1992).
\bibitem{duffy97} D. Duffy, A. Nazarenko, S. Haas, A. Moreo, J. Riera, and E. Dagotto, \prb {\bf 56}, 5597 (1997).
\bibitem{brunner00} M. Brunner, F.~F. Assaad, and A. Muramatsu, \prb {\bf 62}, 15480 (2000).
\bibitem{mishchenko01} A.~S. Mishchenko, N.~V. Prokof'ev, and B.~V. Svistunov, \prb {\bf 64}, 033101 (2001)
\bibitem{hamer98} C.~J. Hamer, W. Zheng, and J. Oitmaa, \prb {\bf 58}, 15508 (1998).

\bibitem{takada03} K. Takada, H. Sakurai, E. Takayama-Muromachi, F. Izumi, R. Dilanian, and T. Sasaki, Nature (London) {\bf 422}, 53 (2003).

\bibitem{azzouz96} M. Azzouz and Th. Dombre, \prb {\bf 53}, 402 (1996).
\bibitem{trumper04} A.~E. Trumper, C.~J. Gazza, and L.~O. Manuel, \prb {\bf 69}, 184407 (2004).

\bibitem{schaak03} R.~E. Schaak, T. Klimezuk, M.~L. T. Foo, R.~J. Cava, Nature {\bf 424}, 527 (2003).
\bibitem{chen04} D.~P. Chen, H.~C. Chen, A. Maljuk, A. Kulakov, H. Zhang, P. Lemmens, C.~T. Lin, \prb {\bf 70}, 24506
2004).
\bibitem{milne04} C.~J. Milne, D.~N. Argyriou, A. Chemseddine, N. Aliouane, J. Veira, S. Landsgesell, D. Alber, \prl {\bf 93}, 247007 (2004).

\bibitem{terasaki03} I. Terasaki, Physica B {\bf 328}, 63 (2003).
\bibitem{lee04}  K.~W. Lee, J. Kunes, W.~E. Picket, \prb {\bf 70}, 045104 (2004).
\bibitem{zheng04} W. Zheng, J. Oitmaa, C.~J. Hamer, and R.~R.~P. Singh, \prb {\bf 70}, 020504 (2004).
\bibitem{watanabe05} H. Watanabe and M. Ogata, J. Phys. Soc. Jpn. {\bf 74}, 2901 (2005).

\bibitem{baskaran03} G. Baskaran, \prl, {\bf 91}, 097003 (2003).

\bibitem{gagliano87} E.~R. Gagliano, and C. Balseiro, Phys. Rev. Lett. {\bf 59}, 2999 (1987).

\bibitem{manousakis91} For a review, see E. Manousakis, \rmp {\bf 63}, 1 (1991).

\bibitem{trugman88} S.~A. Trugman, \prb {\bf 37},  1597  (1988).
\bibitem{white01} S.~R. White and I. Affleck, \prb {\bf 64} 024411 (2001)

\bibitem{starykh95} O.~A. Starykh and G.~F. Reiter, \prb, {\bf 53}, 2517 (1996).
\bibitem{chernyshev99} A.~L. Chernyshev and P.~W. Leung, \prb {\bf 60}, 1592 (1999).

\bibitem{shraiman88} B.~I. Shraiman and E.~D. Siggia, \prl {\bf 60}, 740 (1988).

\end{thebibliography}
\end{document}